\documentclass[conference]{IEEEtran} 
\usepackage{cite}
\usepackage{graphicx}
\usepackage{algorithm}
\usepackage{algorithmic}
\usepackage{listings}
\usepackage{multirow}
\newtheorem{definition}{Definition}

\newtheorem{theorem}{Theorem}

\begin{document}

  \title{An Improved Implementation of Grain}

\author{Shohreh Sharif Mansouri and Elena Dubrova \\
Department of Electronic Systems, School of ICT, KTH - Royal Institute of Technology, Stockholm\\Email:\{shsm,dubrova\}@kth.se}
\maketitle

\begin{abstract}
A common approach to protect confidential information 
is to use a stream cipher which combines plain text bits with a pseudo-random bit sequence. Among the existing 
stream ciphers, Non-Linear Feedback Shift Register (NLFSR)-based ones provide the best trade-off between 
cryptographic security and hardware efficiency. In this paper, we show how to further improve the hardware 
efficiency of Grain stream cipher. By transforming the NLFSR of Grain from its original Fibonacci configuration 
to the Galois configuration and by introducing a clock division block, we double the throughput 
of the 80 and 128-bit key 1bit/cycle architectures of Grain with no area penalty. 
\end{abstract}

\section{Introduction}


Constrained environments applications such as hardware authentication devices (RFID, etc),
smartcards, and wireless networks (Bluetooth, NFC, etc) require power-efficient,
area-efficient and high-performance hardware encryption systems with large security margins.
Until today, no adequate cryptographic solution has been proposed which satisfies the extreme
limitations of devices like RFIDs~\cite{Juels}. Even the most compact of today's encryption systems - Non-Linear
Feedback Shift Register (NLFSR)-based stream ciphers - contain an order of magnitude more gates than can be dedicated
for security functionality in the low-cost RFID tags~\cite{GoB08}. The lack of adequate encryption mechanisms gives
rise to many security and privacy problems and blocks off a variety of potential applications of RFIDs.

Motivated by these needs, in 2004-2008 the EU ECRYPT network carried out the eSTREAM project with the
objective to identify the best stream ciphers designs~\cite{eSTREAM}. Stream ciphers Grain-80~\cite{Hell}, Trivium~\cite{canniere-trivium}, and Mickey-v2~\cite{mickey}
were selected as finalists for the hardware-oriented profile. Grain-80 with 1 bit/cycle throughput has the smallest hardware among all eSTREAM candidates, which makes
it a particularly interesting case. 

Grain-80 consists of one 80-bit LFSR~\cite{LFSR}, one 80-bit NLFSR~\cite{NLFSR}, and a function combining selected state bits. The shift registers take almost 80 percent of the total area of the system and define its critical path. In this paper, we show that by transforming the shift registers of Grain from their original Fibonacci configuration to the Galois configuration, we can significantly improve its throughput.
In the Fibonacci configuration of shift registers, the feedback is applied to the first bit of the register only.
In the Galois configuration, the feedback can be applied to any bit. Thus, the 
depth of the circuits implementing feedback functions of the Galois configuration can potentially be smaller, leading to shorter propagation time and higher
throughput. 

However, unlike the LFSR case in which the mapping from the Fibonacci configuration
to the Galois configuration is one-to-one, in the NLFSR case multiple Galois
NLFSRs can be equivalent to a given Fibonacci one~\cite{Du08j}. The problem of selecting a
"best" Galois NLFSR for a given Fibonacci one is still open. One of the contributions
of this paper is finding the minimal-throughput Galois configurations of NLFSRs for Grain-80 and Grain-128~\cite{grain128}.
Another contribution is the introduction of 
the clock division block which 
divides the clock frequency of Grain by two or four during the initialization phase. 
Without such a block, the potential benefits of the Galois configuration can not be utilized.
\section{Background} \label{back}

\subsection{Definition of NLFSRs}

A {\em Non-Linear Feedback Shift Register (NLFSR)} consists of $n$
binary storage elements, called {\em bits}. 
Each bit $i \in \{0,1,\ldots,n-1\}$ has an associated {\em state variable} $x_i$ 
which represents the current value of the bit $i$ and a {\em feedback function} 
$f_i: \{0,1\}^n \rightarrow \{0,1\}$ which determines how the value of $i$ is updated. 

A {\em state} of an NLFSR is an ordered set of values of its
state variables.
At every clock cycle, 
the next state is determined from the current state 
by updating the values of all bits simultaneously
to the values of the corresponding $f_i$'s.
The {\em output} of an NLFSR is the value of its 0th bit.

If for all  $i \in \{0,1,\ldots,n-2\}$ the feedback functions are of type $f_i = x_{i+1}$, we call
an NLFSR the {\em Fibonacci} type. Otherwise, we call an NLFSR the {\em Galois} type.

Two NLFSRs are {\em equivalent} if their sets of output sequences are equal.


\subsection{The Transformation from the Fibonacci to the Galois Configuration} \label{prel}

Let $f_i$ and $f_j$ be feedback functions of bits $i$ and $j$ of an
$n$-bit NLFSR, respectively. The operation {\em shifting}, denoted by $f_i
\stackrel{P}{\rightarrow} f_j$, moves a set of product-terms $P$
from $f_i$ to $f_j$.
The index of each variable $x_k$ of each product-term in $P$
is changed to $x_{(k-i+j)~\mbox{\small{mod}}~n}$.  

The {\em terminal bit} $\tau$ of an $n$-bit NLFSR is the bit with the maximal index
 which satisfies the following condition: For all bits $i$ such that $i < \tau$, $f_i$ is of type $f_i = x_{i+1}$.

\begin{definition} \label{def_unif}
An $n$-bit NLFSR is {\em uniform} if the following two conditions hold: 
\begin{enumerate}
\item[(a)] all its feedback functions are {\em singular} functions of type
\[
f_i(x_0,\ldots,x_{n-1}) = x_{(i+1) mod~n} \oplus g_i(x_0,\ldots,x_{n-1}),
\]
where $g_i$ does not depend on $x_{(i+1) mod~n}$,
\item[(b)] for all its bits $i$ such that $i > \tau$, 
the index of every variable of $g_i$ is not larger than $\tau$.
\end{enumerate}
\end{definition}

\begin{theorem}~\cite{Du08j} \label{th_main}
Given a uniform NLFSR with the terminal bit $\tau$, a shifting $g_{\tau} \stackrel{P}{\rightarrow} g_{\tau'}$,
 $\tau' < \tau$, results in an equivalent NLFSR if the transformed NLFSR is uniform as well.
\end{theorem}

\section{The Description of Grain} \label{dg}



There are two versions of Grain: 80-bit~\cite{Hell} key and 128-bit key~\cite{grain128}. Both
consist of an LFSR, an NLFSR, and two combining functions. 

In Grain-80 the shift registers are 80-bits. They are both the Fibonacci type, i.e.
all bits except the 79th repeat the value of the previous bit.
The feedback function of the 79th bit of the LFSR 
is given by:
\[
f_{79} = s_{62} \oplus s_{51} \oplus s_{38} \oplus s_{23} \oplus s_{13} \oplus s_{0}
\]
where $s_i$ is the state variable of the $i$th bit, $i \in \{0,1,\ldots,79\}$.

The feedback function of the of the 79th bit of the NLFSR 
is given by:
\[
\begin{array}{l}
g_{79} = s_{0} \oplus b_{0} \oplus b_{62} \oplus b_{60} \oplus b_{52} \oplus b_{45} \oplus b_{37} \oplus b_{33} \oplus b_{28}\\
 \oplus b_{21} \oplus b_{14} \oplus b_{9}\oplus b_{63} b_{60} \oplus b_{37} b_{33} \oplus b_{15} b_{9} \oplus b_{60} b_{52} b_{45}\\
   \oplus b_{33} b_{28} b_{21} \oplus b_{63} b_{45} b_{28} b_{9} \oplus b_{60} b_{52} b_{37} b_{33}  \oplus b_{63} b_{60} b_{21} b_{15}\\
     \oplus b_{63} b_{60} b_{52} b_{45} b_{37} \oplus  b_{33} b_{28} b_{21} b_{15} b_{9} \oplus   b_{52} b_{45} b_{37} b_{33} b_{28} b_{21}\\
\end{array}
\]
where $b_i$ is the state variable of the $i$th bit, $i \in \{0,1,\ldots,79\}$.

The first combining function of Grain-80 produces it output value based of the selected bits 
from the NLFSR and the LFSR:
\[
\begin{array}{l}
H = s_{25} \oplus b_{63} \oplus s_3 s_{4} \oplus s_{46} s_{4} \oplus s_{4} b_{63} \oplus s_3 s_{25} s_{46} 
       \oplus \\
 s_3 s_{46} s_{4} \oplus \index{\footnote{}} s_3 s_{46} b_{63} \oplus s_{25} s_{46} b_{63} \oplus s_{46} s_{4} b_{63}
\end{array}
\]
The second combining function of Grain-80 generates the output stream of the system
from the selected bits from the NLFSR and LFSR states and the output of $H$:
\[
Z = \sum_{k \in A} b_{k} \oplus H , 
\]
where $A = \{1, 2, 4, 10, 31, 43, 56\}$.

For Grain-128, the corresponding functions are:
\[
\begin{array}{l}
~~f_{127} = s_{0} \oplus s_{7} \oplus s_{38} \oplus s_{70} \oplus s_{81} \oplus s_{96} \\[2mm]
\begin{array}{l}
g_{127} =  s_{0} \oplus b_{0} \oplus b_{26} \oplus b_{56} \oplus b_{91} \oplus b_{96}\oplus b_{3} b_{67} \oplus b_{11} b_{13}   \\
           \oplus b_{17} b_{18} \oplus    \oplus b_{27} b_{59} \oplus b_{40} b_{48}  \oplus   b_{61} b_{65}\oplus   b_{68} b_{84} \\[2mm]
\end{array} \\
~~H = b_{12} s_8 \oplus s_{13} s_{20} \oplus b_{95} s_{42}  \oplus s_{60} s_{79} \oplus b_{12} b_{95} s_{95}\\[2mm]
~~Z = \sum_{k \in A} b_{k} \oplus s_{93} \oplus H 
\end{array}
\]
where $A = \{2, 15, 36,  45, 64, 73, 89\}$.


Before generating a stream of data, a cipher must be initialized with default keys. During the initializing phase the cipher does not produce any 
output for 160 clock cycles for Grain-80 and 256 cycles for Grain-128. The output of the $Z$ function is XOR-ed with the outputs of LFSR and NLFSR 
and then fed into the inputs of both shift registers, as shown in Figure \ref{fig:longest-path}. After the initialization, the loops are opened and there is no feedback between the two shift registers.



It is possible to increase the throughput of Grain at the expense
 of extra hardware by introducing parallelism in its architecture. In parallelized versions of Grain, in each
clock cycle blocks of duplicated NLFSR and LFSR feedback functions produce output bits in parallel.
To allow for up to 16 (32) degrees of parallelization, Grain-80 (128) is designed
so that the bits $65 < i < 79$  ($97 < i < 127$) of the shift registers are not used in
the feedback functions or in the input to the combining functions.


\section{Grain with Galois Configuration} \label{gg}

Grain can be modified by transforming its LFSR and NLFSR from their original Fibonacci configurations
to the Galois configurations. The transformation of LFSRs is done using standard techniques, 
in this section we only describe the transformation of NLFSRs.

The NLFSR of Grain-80 (128) can be transformed to the Galois configuration by shifting the product-terms
of the feedback function of 79th (127th bit) to the feedback functions
of bits with lower indexes. By Theorem~\ref{th_main}, if the NLFSR after shifting 
 satisfies the conditions of the Definition~\ref{def_unif}, then it produces  
the same sets of output sequences as the NLFSR before shifting.

Ideally, in order to maximize the throughput, we want to distribute the products equally among feedback functions. 
However, according to~\cite{Du08j},
to guarantee equivalence of NLFSRs before and after shifting, we cannot shift to bits with indexes lower that the bit $\tau$
which is given by:
\[
\begin{array}{lccl}
\tau & = & max & (max\_index(p)-min\_index(p)), \\
     & & \forall p \in P & 
\end{array}
\]
where $P$ is the set of all product-terms of the feedback function of the Fibonacci NLFSR, and $min\_index(p)$
($max\_index(p)$) denotes the minimal (maximal) index of variables the product-term $p$.

For Grain-80, the product-term with the maximal difference in indexes of variables is $b_{63} b_{45} b_{28} b_{17} b_9$,
so $\tau = 54$. For Grain-128, we have $\tau = 64$ due to the product-term  $b_3 b_{67}$.

However, in order to avoid modifications of the encrypting algorithm of Grain, we need to guarantee not only the equivalence of the sequences of output bits, but
also the equivalence of the sequences of of all internal bits of the NLFSR used by the
combining functions. A modification of the encrypting algorithm could lead to undesirable changes in the Grain security. For Grain-80, the bit 63 of the NLFSR is used in the function $H$,
and bits $1, 2, 4, 10, 31, 43, 56$ are used in the function $Z$. Since 56 and 63 are greater than 54, we cannot 
use $\tau = 54$ as the terminal bit of the Galois configuration. We need to set the terminal bit to 63.
Then, for all bits $i \in \{0,1,\ldots,62\}$, the 
feedback functions will be of type $g_i = b_{i+1}$, an the output sequences of the bits $i \in \{1,2,\ldots,63\}$
will be the same as the output sequence of the bit 0 shifted in time. Consequently, the algorithm of Grain
will not change.

For Grain-128, the bits 12 and 95 of the NLFSR are used in $H$ and the bits $2, 15, 36,  45, 64, 73, 89$ are used
in $Z$. Therefore, the terminal bit has to be 95.

After we have chosen the position of the terminal bit, we can start shifting products from
the function $g_{79} (g_{127})$ to the functions with indexes larger or equal than the
terminal bit. Shifting can be done in many different ways. At present there is no systematic
technique which guarantees that the transformation produces an NLFSR with
the minimal throughput for a given technology. We found the solutions presented 
below by trying many different choices.

\subsection{One Bit per Cycle Version}

According to our simulation results, the following Galois NLFSR results in the maximal throughput for 1bit/cycle version of Grain-80 :
\[
\begin{array}{l}
g_{79} =  s_{0} \oplus b_{0} \oplus b_{37}   \\
g_{78} =  b_{79} \oplus  b_{44} \\
g_{77} =  b_{78} \oplus  b_{50} \\
g_{76} =  b_{77} \oplus  b_{57} \\
g_{75} =  b_{76} \oplus  b_{58} \\
g_{74} =  b_{75} \oplus  b_{32} b_{28} \\
g_{73} =  b_{74} \oplus  b_{3} \\
g_{72} =  b_{73} \oplus  b_{8} b_{2} \\
g_{71} =  b_{72} \oplus  b_{55} b_{37} b_{20} b_{1} \\
g_{70} =  b_{71} \oplus  b_{24} b_{19} b_{12} b_{6} b_{0} \\
g_{69} =  b_{70} \oplus  b_{53} b_{50} \\
g_{68} =  b_{69} \oplus  b_{49} b_{41} b_{26} b_{22} \\
g_{67} =  b_{68} \oplus  b_{9} \oplus  b_{21} b_{16} b_{9} \\
g_{66} =  b_{67} \oplus  b_{15} \oplus  b_{47} b_{39} b_{32} \\
g_{65} =  b_{66} \oplus  b_{0} \oplus  b_{38} b_{31} b_{23} b_{19} b_{14} b_{7} \\
g_{64} =  b_{65} \oplus  b_{18} \oplus  b_{48} b_{45} b_{6} b_{0} \\
g_{63} =  b_{64} \oplus  b_{47} b_{44} b_{36} b_{29} b_{21}
\end{array}
\]
Here and further in this section, all omitted feedback functions are of type $g_{i} =  b_{i+1}$.

For Grain-128, the maximal-throughput Galois NLFSR is:
\[
\begin{array}{l}
g_{127} =  s_{0} \oplus b_{0} \oplus  b_{3} b_{67} \\
g_{124} =  b_{125} \oplus  b_{0} b_{64} \\
g_{116} =  b_{117} \oplus  b_{0} b_{2} \\
g_{110} =  b_{111} \oplus  b_{0} b_{1} \\
g_{102} =  b_{103} \oplus  b_{71} \\
g_{101} =  b_{102} \oplus  b_{0} \\
g_{100} =  b_{101} \oplus  b_{0} b_{32} \\
g_{99} =  b_{100} \oplus  b_{63} \\
g_{98} =  b_{99} \oplus  b_{27} \\
g_{97} =  b_{98} \oplus  b_{38} b_{54} \\
g_{96} =  b_{97} \oplus  b_{30} b_{34} \\
g_{95} =  b_{96} \oplus  b_{8} b_{16} 
\end{array}
\]

\subsection{Multiple Bit per Cycle Version}

We can extend the theory presented in~\cite{Du08j},~\cite{elena2} to $k$ bits/cycle versions of Grain by  
restricting bit positions to which the feedback can be applied. 
It is easy to see that, to ensure times $k$ degree of parallelization of an $n$-bit Galois NLFSR with the 
terminal bit $\tau$, all bits except 
\[
n-1- i \cdot k, ~ \mbox{for all} ~ i = \{0,1,\ldots, \lfloor (n-1-\tau)/k \rfloor - 1 \}
\] 
should have feedback functions of type $g = b_{(i + 1)}$.

So, for example, for 4 bits/cycle version of Grain-80, we can apply feedback to the bits 79,75,71 and 67:
\[
\begin{array}{l}
g_{79} = s_{0} \oplus  b_{0} \oplus  b_{62} \oplus  b_{33} \oplus  b_{28} \oplus  b_{21} \oplus b_{15} b_{9} \\
~~~~~~~ \oplus b_{52} b_{45} b_{37} b_{33} b_{28} b_{21} \\
g_{75} = b_{76} \oplus  b_{41} \oplus  b_{33} \oplus  b_{5} \oplus  b_{59} b_{56} \oplus  b_{33} b_{29} \oplus  b_{59} b_{41} b_{24} b_{5} \\
g_{71} = b_{72} \oplus  b_{44} \oplus  b_{25} b_{20} b_{13} \oplus  b_{55} b_{52} b_{13} b_{7} \oplus  b_{25} b_{20} b_{13} b_{7} b_{1} \\
g_{67} = b_{68} \oplus  b_{48} \oplus  b_{2} \oplus  b_{48} b_{40} b_{33} \oplus  b_{48} b_{40} b_{25} b_{21} \\
~~~~~~~	\oplus  b_{51} b_{48} b_{40} b_{33} b_{25} 
\end{array}
\]
For 8 bit/cycle version of Grain-80, we can apply feedback to the bits 79 and 71:
\[
\begin{array}{l}
g_{79} =  s_{0} \oplus  b_{0}  \oplus  b_{14} \oplus  b_{9} \oplus  b_{15} b_{9} \oplus  b_{60} b_{52} b_{45} \oplus  b_{33} b_{28} b_{21} \\
 ~~~~~~~   \oplus b_{60} \oplus  b_{60} b_{52} b_{37} b_{33} \oplus  b_{63} b_{60} b_{21} b_{15} \oplus  b_{33} b_{28} b_{21} b_{15} b_{9} \\
g_{71} = b_{72}  \oplus  b_{44} \oplus  b_{37} \oplus  b_{29} \oplus  b_{25} \oplus  b_{20} \oplus  b_{13} \oplus  b_{55} b_{52} \\
~~~~~~~ \oplus  b_{54} \oplus  b_{29} b_{25} \oplus  b_{55} b_{37} b_{20} b_{1} \oplus  b_{55} b_{52} b_{44} b_{37} b_{29}  \\
~~~~~~~ \oplus  b_{44} b_{37} b_{29} b_{25} b_{20} b_{13}
\end{array}
\]
For 16 bit/cycle version of Grain-80, we can apply feedback only to the bit 79, i.e. no transformations can be done.

For 4 bit/cycle version of Grain-128, we can apply feedback to the bits 127,123,119,115,111,107 103 and 99:
\[
\begin{array}{l}
g_{127} = s_{0} \oplus b_{0} \oplus  b_{3} b_{67} \\
g_{123} = b_{124} \oplus  b_{64} b_{80} \\
g_{119} = b_{120} \oplus  b_{3} b_{5} \\
g_{115} = b_{116} \oplus  b_{49} b_{53} \\
g_{111} = b_{112} \oplus  b_{1} b_{2} \\
g_{107} = b_{108} \oplus  b_{6} \oplus  b_{76} \\
g_{103} = b_{104} \oplus  b_{67} \oplus  b_{3} b_{35} \\
g_{99} = b_{100} \oplus  b_{28} \oplus  b_{12} b_{20} \\
\end{array}
\]
For 8 bit/cycle version of Grain-128, we can apply feedback to the bits 127,119,111,103:
\[
\begin{array}{l}
g_{127} = s_{0} \oplus b_{0} \oplus  b_{56} \oplus  b_{3} b_{67} \\
g_{119} = b_{120} \oplus  b_{18} \oplus  b_{88} \oplus  b_{3} b_{5} \\
g_{111} = b_{112} \oplus  b_{75} \oplus  b_{1} b_{2} \oplus  b_{52} b_{68} \\
g_{103} = b_{104} \oplus  b_{3} b_{35} \oplus  b_{16} b_{24} \oplus  b_{37} b_{41} \\
\end{array}
\]
For 16 bit/cycle version of Grain-128, we can apply feedback to the bits 127 and 111.
\[
\begin{array}{l}
g_{127} = s_{0} \oplus b_0 \oplus b_{56} \oplus b_3 b_{67} \oplus b_{11} b_{13} \oplus b_{40} b_{48} \\
g_{111} = b_{112} \oplus b_{10} \oplus b_{75} \oplus b_{80} \oplus b_1 b_2 \oplus b_{11} b_{43} \oplus b_{45} b_{49} \\
 ~~~~~~~    \oplus b_{52} b_{68}
\end{array}
\]
  For 32 bit/cycle version of Grain-128, we can apply feed-back only to the bit 127, i.e. no transformations can be done.

\subsection {Design Details}

\begin{figure}[htbp]
	\centering
		\includegraphics[scale=0.37]{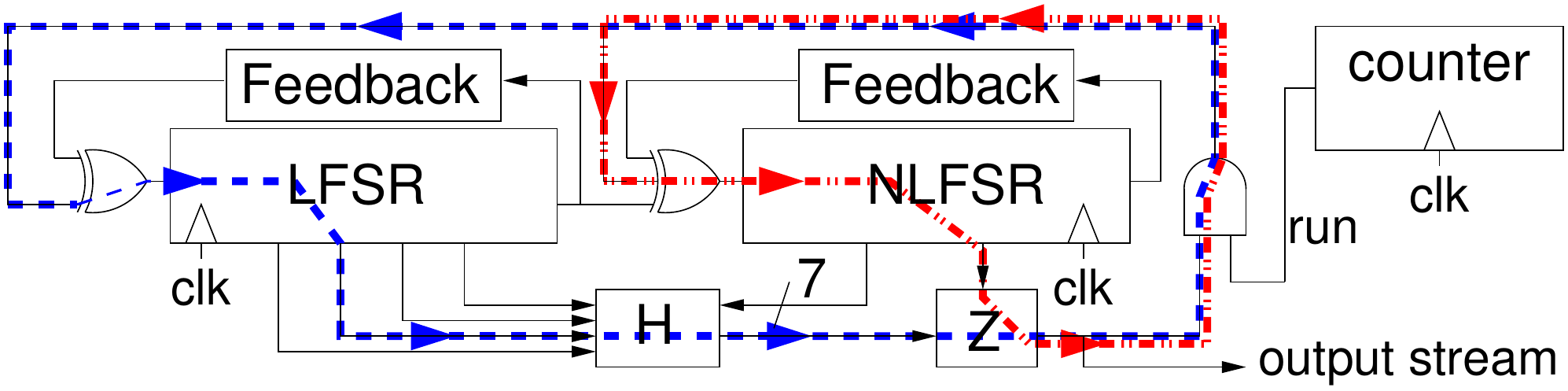}
	\caption{Initialization loops Grain}
        \label{fig:longest-path}
\end{figure}

By transforming Grain's shift registers to the Galois configuration as described in the previous section, 
we can obtain up to 40 \%
reduction in their propagation time. However, the clock frequency 
of the overall Grain system improves only 
about 10\%. The problem is in the hardware architecture of Grain during key
initialization, during which the output value of $Z(x)$ is fed back to the LFSR and
NLFSR making two loops, as shown in Figure \ref{fig:longest-path}. 
After the
transformation from the Fibonacci to the Galois configuration, due to the reduction of the
critical path in the NLFSR, the critical path of the system is no longer in the
NLFSR but is in the initialization loops, which are closed only during initialization.
Thus, the highest frequency that the system supports during initialization
is lower than the highest frequency supported during key stream generation (see Table~\ref{table:inital freq}).
\vspace{-1.0em}
\begin{table}[htbp]
\begin{center}
\caption{Clock frequencies of Galois Grain-80}
    \begin{tabular}{ | c| c c  |  }  \hline

\multicolumn{1}{|c|}{Galois Grain-80}& \multicolumn{2}{c|}{Maximum Clock Frequency } \\  \cline{2-3}
\multicolumn{1}{|c|}{Block Size}& Initialization&  Key Generation  \\  
\hline\hline
 1 bit/cycle& 1 (GHz) & 4 (GHz)\\  \hline
 4 bits/cycle& 1 (GHz)& 2.7 (GHz) \\  \hline
 8 bits/cycle& 1 (GHz)& 2.3 (GHz) \\  \hline
      \end{tabular}
\label{table:inital freq}
\end{center}
\end{table}
\vspace{-1.0em}
\begin{figure}[htbp]
	\centering
		\includegraphics[scale=0.37]{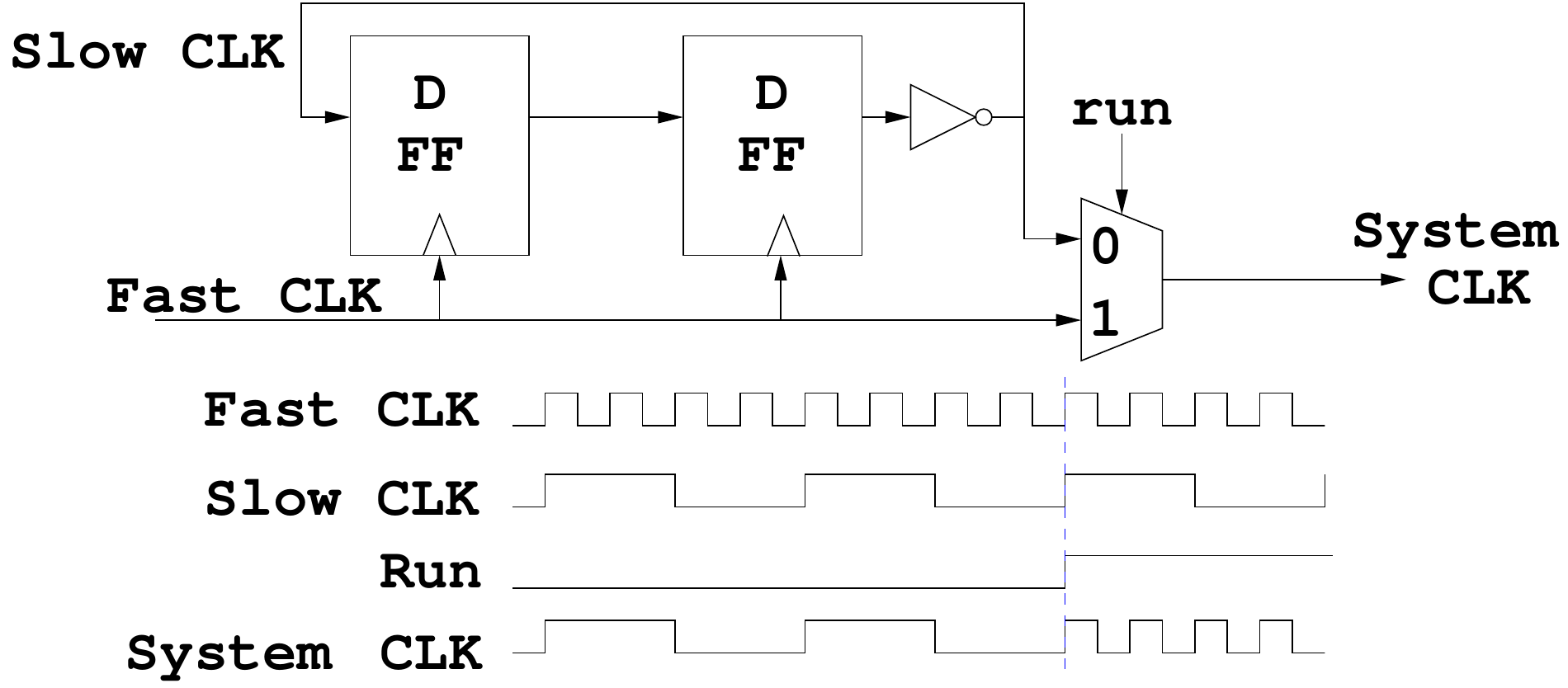}
	\caption{Clock divider by four}
        \label{fig:double_clk4}
\end{figure}

\begin{table*}[htbp]
\begin{center}
\caption{Synthesis results for Galois and Fibonacci configurations of Grain
in TSMC 90nm technology} 
    \begin{tabular}{| l | c | l  l  l | l  l  l |l  l  l | l  l  l | }  
  \hline

 \multicolumn{1}{|c|}{Cipher}&\multicolumn{1}{c|}{Block Size}& \multicolumn{3}{c|}{Frequency (GHz)}& \multicolumn{3}{c|}{Area (GE)}&
\multicolumn{3}{c|}{Power (mW)}& \multicolumn{3}{c|}{Throughput (Gbit/Sec)}\\
\cline{3-14}\cline{3-14}
      && Galois & Fib. & impr. & Galois & Fib. & impr.  & Galois & Fib.& impr. & Galois & Fib. & impr. 
\\\cline{1-14}
     \hline \hline

  & 1  & 4   & 1.9 & 110\% & 1772 &1743   & 0\%  &  0.035  & 0.056  & 37\% & 4 & 1.9 &  110\% \\  \cline{2-14}
 Grain-80 & 4  &  2.7  & 1.8 & 50\%    & 2471  & 2589  & 5\%    & 0.095  & 0.099 & 4\%  & 10.8  &  7.2 & 50\%    \\  \cline{2-14}
  & 8   &2.3& 1.8  & 28\%   & 3575   &  3867   & 8\% &0.127& 0.149  & 15\% &18.4 & 14.4    & 28\% \\  \hline

& 1  & 4.6 &  2 & 130\%& 2207 & 2408  & 8\% &0.12  & 0.17 &29\% &4.6 &  2 &130\%\\  \cline{2-14}
Grain-128&  4  &  3.5  & 2 &75\%   &3021  &  3182 &5\% & 0.13  & 0.14 & 7\%  &14&8 & 75\%   \\ \cline{2-14}
&  8   & 3.2&2  & 60\% & 3902 &    4143 & 6\%  & 0.15& 0.17  &12\% &25.6  & 16&60\% \\ \cline{2-14}
&  16  &2.7&2 &35\%  &5346&5474 &2\% &0.20& 0.21&  5\% &43.2&32 &35\%\\ \hline

    \end{tabular}
\label{table:grain801}
\end{center}

\end{table*}

To obtain a higher improvement in the throughput of Grain, we introduce a clock division block to divide the frequency of the clock during
the initialization phase. The clock divider is realized as a simple block containing
one or two D-flipflops which divides the clock frequency of the system by 2 or 4. In
Figure \ref{fig:double_clk4} we show the structure of the clock division block for division of the clock frequency by 4.
In some versions of Grain, division by 2 is sufficient to ensure correct operation during the initialization phase.
Division by 3 would be suitable in some cases, but it would overcomplicate the hardware for only a modest speedup of the initialization phase.
The clock division block is a very small component. Clock division by four gives area overhead of 25.67 GE and negligible power consumption.
\begin{figure}[htbp]
	\centering
		\includegraphics[scale=0.37]{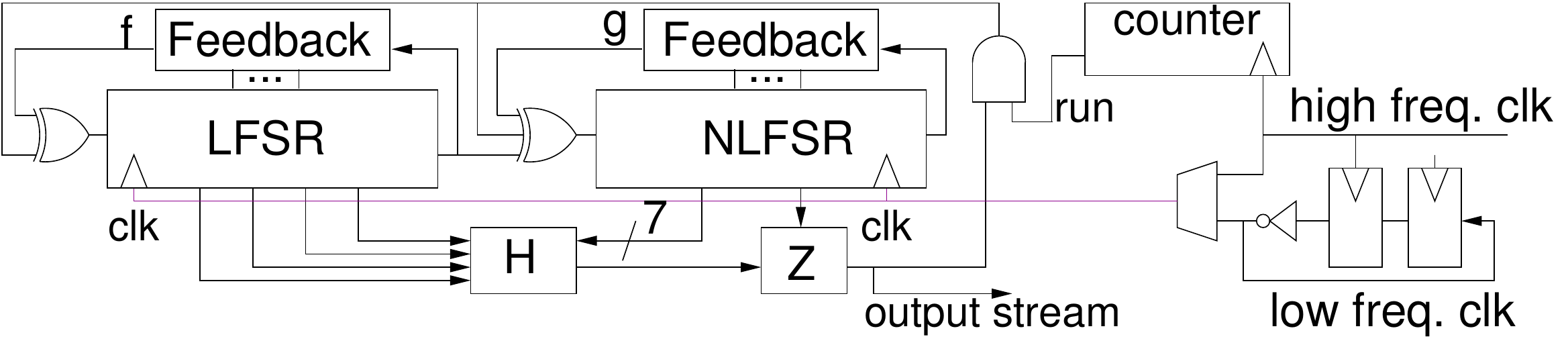}
	\caption{Grain with Galois configuration and clock divider
by four}
        \label{fig:new-grain-sdf}
\end{figure}

\begin{figure}[htbp]
\centering
\includegraphics[scale=0.37]{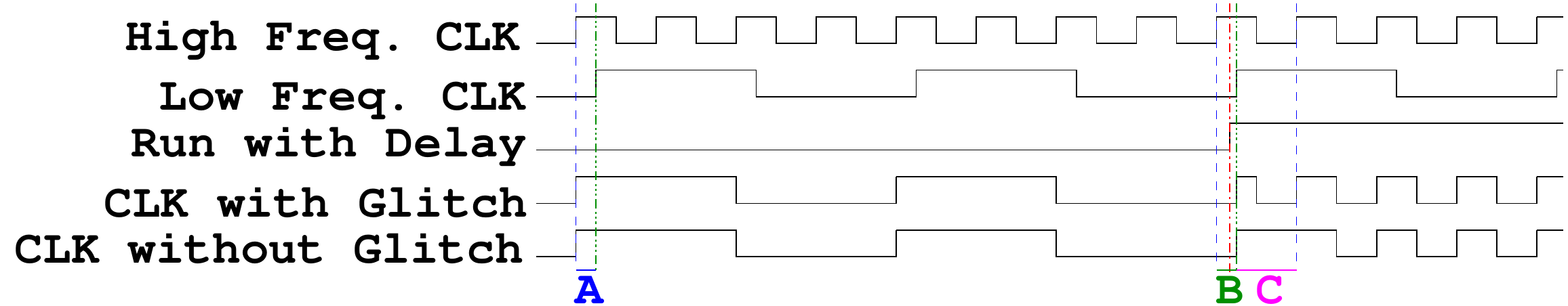}
\caption{A: delay between positive edges of two clocks. B:delay between run signal
and high frequency clock. C: shortened clock}
        \label{fig:clkglitch1}
\end{figure}

\vspace{-1.0em}
Grain always moves from the slower to the faster clock frequency and the run
signal is set internally by the counter on the positive edge of the clock.
Because of the delay in the production of the run signal, the first clock cycle
of the key generation phase will be shortened, which could potentially lead to
critical path violations in a performance-optimized design such as Grain. We can
handle this problem by using a flip-flop in front of the run signal which is
output by the counter. In this case, 
if the run signal rises to 1 after a positive edge of the faster clock signal,
the clock of the system changes to the faster clock in the next positive edge of
the system. This solution is shown in Figure \ref{fig:clkglitch1}.

\section{Experimental Results}\label{r}

We have synthesized the Fibonacci and the Galois versions of Grain using Cadence RTL compiler 
in the TSMC 90nm standard cell technology library. Since the synthesis tool does not handle
multiple clocking, we set the two initialization loops as false paths and optimized 
the designs for the key-generation phase.

Table~\ref{table:grain801} shows the results for throughput, power consumption, area, and frequency. 
Area is measured in terms of NAND2 Gate Equivalents (GE). 
The total power consumption of the system is estimated as a combination of dynamic
and leakage power for operation at 25 C, with a power supply of 1.2
V at 10MHz clock frequency as in ~\cite{GoB08}.

As we can see, the throughput for Galois 1bit/cycle Grain-80 and Grain-128 is more than doubled compared to Fibonacci.

\begin{figure}[htbp]
\centering
\includegraphics[scale=0.37]{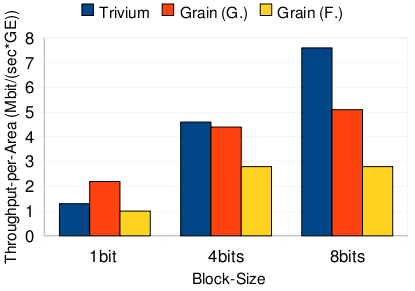}
\caption{The Throughput/area ratios of Trivium and Grain}
        \label{fig:throughput}
\end{figure}

Trivium is the highest ranked finalist in the eSTREAM project. In Table
\ref{table:triviumGrain}, we compared the frequency and area of Trivium (T) and Grain-80 with
Galois configuration(G). Both ciphers were implemented in TSMC 90 nm technology.
Due to the Galois configuration, Grain-80 (1bit/cycle) is faster and
smaller than Trivium (1bit/cycle), with a significantly better throughput/area ratio.
This is an important result for applications such as RFID systems which require efficiency in both throughput and area.
The throughput/area figures are compared graphically in Figure~\ref{fig:throughput}, where the figures
for the Fibonacci configuration (Grain(F)) of Grain-80 are also reported. 
\begin{table}[htbp]
\begin{center}
\caption{Comparison between Trivium and Grain-80} 
    \begin{tabular}{ | c | l  l   | l  l   |}  \hline

 \multicolumn{1}{|c|}{Block}& \multicolumn{2}{c|}{Freq (GHz)}& \multicolumn{2}{c|}{Area (GE)}\\\cline{2-5}

 \multicolumn{1}{|c|}{Size}&\multicolumn{1}{|c}{T}&\multicolumn{1}{c|}{G}&\multicolumn{1}{|c}{T}&\multicolumn{1}{c|}{G}   \\
     \hline\hline
  1  & 3.8   &4&    2810 & 1772   \\  \hline
  4  & 3.4  & 2.7&   2955  & 2471     \\  \hline
  8  & 3.6 & 2.3&   3763 &3575    \\  \hline
  16 & 3.6 & 1.7&   3841  & 5768    \\  \hline
    \end{tabular}


\label{table:triviumGrain}
\end{center}
\end{table}
\vspace{-2.0em}

\section{Conclusion}\label{c}

In this paper, we presented an improved version of the Grain stream cipher.
We found new implementations for its NLFSRs which
generate the same cryptographically strong pseudo-random bit sequences
as the ones of the original Grain, but have a better hardware efficiency.
The presented technique is general and 
can be applied to any NLFSR-based stream cipher. Its efficiency depends on the feedback 
ufnction of the NLFSR and the desired degree of parallelization. For Trivium the presented technique brings no improvement.





\end{document}